# MEMS Fiber-Tip Photoacoustic Spectrometer for In Situ Microscale Trace Gas Sensing

Xingjie Liu[1,†], Mengqi Wu[1,†], Ping Lu[1,2,*]

[1]National Engineering Research Center of Next Generation Internet Access System, School of Optical and Electronic Information, Huazhong University of Science and Technology, Wuhan 430074, China

[2]Optoelectronic Valley Laboratory, Wuhan 430074, China

## ABSTRACT

To meet the stringent requirements for miniaturized and highly sensitive trace gas sensing in space-constrained scenarios, including power equipment monitoring, minimally invasive biomedical diagnostics, and in situ lithium-battery analysis, we report a MEMS-integrated fiber-tip photoacoustic spectrometer (MFPAS). The device incorporates a Fabry–Pérot (F-P) photoacoustic sensor formed by directly butt-coupling a single-mode fiber (SMF) to a 3 mm×3 mm MEMS chip with a 100-nm-thick low-pressure chemical vapor deposition (LPCVD) $Si_3N_4$ diaphragm. The resulting approximately 200μm deep silicon microcavity functions simultaneously as a photoacoustic gas cell and an acoustic confinement cavity. A micro-aperture fabricated at the diaphragm periphery by focused ion beam (FIB) milling serves as both a gas diffusion channel and an acoustic high-pass filter, suppressing ambient low-frequency pressure fluctuations and stabilizing the F-P quadrature point without active servo control. In gas-phase measurements, the sensor achieves a noise-equivalent concentration (NEC) of 58.5 ppb@1s, with a rapid response time of 6 s. Benefiting from its ultra-small cavity volume of approximately 1.5 nL, the device is further adapted through structural packaging for in situ dissolved gas analysis in transformer oil, where it achieves an NEC of 230 ppb@1s and a T90 response time of 320 s in the oil phase. By combining nanoliter-scale detection volume, ppb-level sensitivity, rapid response, and wafer-scale batch fabrication compatibility, the proposed MFPAS bridges MEMS diaphragm micromachining and FIB-enabled gas exchange engineering. This design overcomes the intrinsic gas-exchange limitation of conventional sealed-diaphragm optical microphones and offers significant potential for power equipment monitoring and in situ health diagnostics.

**Keywords:** Photoacoustic spectroscopy; Fiber-tip Fabry–Pérot sensor; MEMS silicon nitride diaphragm; In situ dissolved gas analysis

## 1 Introduction

The application demand for miniaturized, highly sensitive trace gas sensors is continuously growing in fields such as power equipment status monitoring[1], minimally invasive biomedical detection[2], and in situ health diagnostics of lithium-ion batteries [3]. These scenarios typically require sensors to possess ppb-level detection sensitivity, rapid temporal response, compact footprints suitable for narrow or confined spaces, and extremely low sample consumption. Laser spectroscopic techniques, including cavity ring-down spectroscopy (CRDS) [4], tunable diode laser absorption spectroscopy (TDLAS) [5], photothermal spectroscopy (PTS) [6,7], and photoacoustic spectroscopy (PAS) [8], have demonstrated exceptional performance in trace gas detection. However, conventional schemes generally rely on absorption paths of tens of centimeters to several meters, or centimeter-scale resonant acoustic cavities to accumulate sufficient signal strength [5,9], rendering further miniaturization of the device volume highly challenging.

In recent years, fiber-tip Fabry-Pérot (F-P) microcavities have emerged as a frontier in miniature spectroscopic gas sensor research due to their capability to simultaneously integrate a gas cell, an acoustic cavity, and optical interference within a sub-nanoliter volume [10]. Nevertheless, the intrinsic trade-off between the absorption path length and the interference contrast hinders further sensitivity improvements. Although increasing the cavity length accumulates more photothermal phase modulation, beam divergence subsequently degrades the interference fringe contrast; this mutual constraint severely affects sensing performance [11,12]. To address this, Zhao et al. [13] 3D-printed an open, highly reflective microcavity on a single-mode fiber (SMF) end-face via two-photon polymerization to enhance the contrast, achieving a noise equivalent concentration (NEC) of 160 ppb for $C_2H_2$ at 433 s with a response time of <0.5 s. Yan et al. [14] fabricated focusing spherical surfaces coated with high-reflection films on two SMF end-faces to construct a high-finesse cavity, thereby lowering the NEC to 10 ppb at 440 s. While these external structures effectively enhance sensitivity, their fabrication involves complex processes such as two-photon polymerization or precision coating, limiting batch mass production. Zhao et al. [15] introduced a dual-mode Vernier effect within a single intrinsic F-P cavity, achieving an NEC of 12 ppb for $C_2H_2$ at 303 s using a 1-mm-long Bragg hollow-core fiber F-P cavity; nonetheless, the fundamental constraint between the short-cavity photothermal signal and the interference contrast persists. Distinct from the aforementioned photothermal schemes, Ma et al. [2] utilized confined acoustic pressure enhancement to compensate for the short-cavity loss, achieving a $C_2H_2$ NEC of ~9 ppb at 200 s and an 18 ms response time with a 125-μm-diameter, 60-μm-long fiber-tip F-P cavity. Furthermore, Li et al. [16] monolithically integrated a PSS-OMR and a micro-PAC on an SMF end-face, leveraging a 105 kHz mechanical resonance to enhance the signal-to-noise ratio (SNR) by ~5.5 times. This yielded an NEC of 55 ppb at 356 s and a response time of <0.2 s without an external PAC. Both works demonstrate the inherent advantages of photoacoustic schemes in miniaturization. Despite these advancements significantly driving the development of fiber-optic miniature gas sensors, common challenges remain. Regarding manufacturing, processes like two-photon polymerization and precision coating suffer from poor consistency; sub-micron films [2] are susceptible to pinhole defects; and 3D-printed microbeams [13] are constrained by current lithographic resolutions. In terms of gas channels, femtosecond laser sidewall drilling [11,17] compromises cladding integrity, whereas gas diffusion through splice seams [15] relies heavily on fabrication parameters and is difficult to reproduce. Consequently, existing approaches struggle to simultaneously ensure transducer performance, fabrication consistency, and practicality.

MEMS-fabricated $Si_3N_4$ transducer diaphragms have been widely utilized in fiber-optic F-P pressure [18] and acoustic sensing [19]. The wafer-scale low-pressure chemical vapor deposition (LPCVD) process endows the diaphragms with nanoscale thickness uniformity and highly consistent residual stress, ensuring morphological integrity and excellent device-to-device consistency. Furthermore, due to the inherent high tensile stress of the $Si_3N_4$ diaphragm, the mechanical behavior of the nanometer-thick film is dominated by tensile stress rather than bending stiffness. Consequently, the acoustic sensitivity scales quadratically with the diaphragm radius ($a^2$) rather than following the $a^4$ dependence of a standard plate, enabling a robust acoustic response even at small diaphragm diameters. However, MEMS-fabricated diaphragms are typically fully sealed structures. Introducing a gas channel for photoacoustic gas detection without compromising the diaphragm's acoustic performance poses a formidable challenge. Focused ion beam (FIB) milling offers nanoscale spatial positioning accuracy and is widely used in semiconductor cross-sectioning and nanostructure trimming [20]. Its precise confinement of the processing area makes localized material removal on suspended diaphragms feasible. Therefore, this work proposes FIB-based micro-pore milling to create gas diffusion channels at the diaphragm periphery in a controlled manner, concurrently maintaining the morphological integrity of the main vibratory region.

This paper reports a MEMS-integrated fiber-tip photoacoustic spectrometer. A MEMS chip bearing a LPCVD $Si_3N_4$ diaphragm is directly butt-coupled to a SMF to form a monolithic F-P photoacoustic sensor. The 3 mm × 3 mm chip features a 100-nm-thick $Si_3N_4$ diaphragm grown via LPCVD. The ~200-μm-deep silicon microcavity, formed between the diaphragm and the SMF end-face, functions simultaneously as a photoacoustic gas cell and an acoustic confinement cavity. FIB milling is employed to fabricate a micro-pore at the diaphragm periphery, which acts concurrently as a gas diffusion channel and an acoustic high-pass filter. The latter suppresses low-frequency ambient pressure fluctuations, stabilizing the F-P quadrature point without requiring active servo control. In the gas-phase measurement, this sensor achieved an NEC of 58.5 ppb within a 1-second integration time, and 27 ppb NEC within a 103-second integration time. The response time was 6 seconds, and the cavity volume was approximately 1.5 nL. Following adaptive structural packaging, the device is applicable for in situ dissolved gas analysis in power transformers. In the oil-phase environment, the sensor yields an NEC of 206.6 ppb at a 1-s integration time with a T90 response time of 320 s. Combining the advantages of a sub-nanoliter detection volume, a ppb-level detection limit, rapid response, and wafer-scale fabrication consistency, the proposed sensor exhibits substantial engineering application potential in micro-scale detection fields such as online status monitoring of power equipment and in situ health diagnostics.

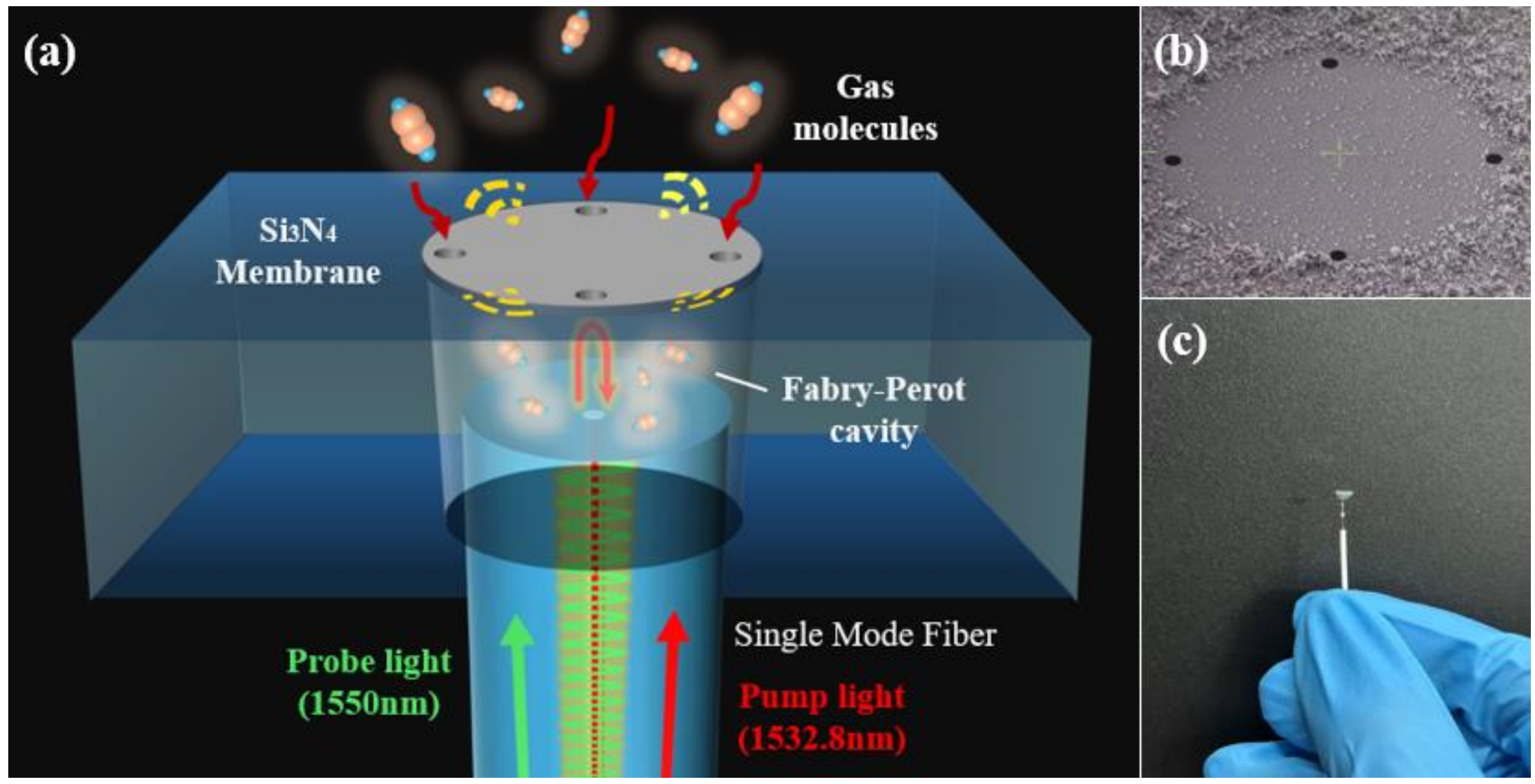

Fig. 1 (a) Three-dimensional structure. (b) Scanning electron microscope (SEM) image of the diaphragm after FIB micro-pore milling. (c) Photograph of the overall MFPAS structure

## 2 Principle and Analysis

### 2.1 Sensor Structure

The three-dimensional structure of the proposed MEMS-integrated fiber photoacoustic spectrometer (MFPAS) is schematically illustrated in Fig. 1(a). The sensor consists of a SMF and a silicon-based MEMS chip, combined to form a monolithic probe structure. A 128-μm through-channel is fabricated on a 3 mm × 3 mm silicon substrate using a deep reactive ion etching (DRIE) process for the precise insertion and fixation of the fiber. A 100-nm-thick $Si_3N_4$ diaphragm is grown on the front side of the chip via LPCVD, forming a free-standing, suspended diaphragm region with an effective diameter of 128 μm centered at the channel orifice. This diaphragm region simultaneously serves as a reflective surface, constituting a F-P interference cavity with the SMF end-face. Several micro-pores are milled around the effective area of the diaphragm via a FIB process to serve as channels for gas entering and exiting the cavity. Both the pump and probe beams are transmitted to the sensing probe through the same SMF. The modulated pump beam is delivered into the F-P cavity via the SMF, and target gas molecules absorb the pump photon energy, releasing heat through non-radiative relaxation. This process generates periodic pressure fluctuations within the sealed cavity, namely the photoacoustic (PA) wave. Since the surrounding silicon walls of the $Si_3N_4$ diaphragm constitute a rigid acoustic boundary, a localized acoustic pressure enhancement effect is established within the cavity, confining the PA wave energy inside the cavity rather than diffusing it into free space. The localized enhanced acoustic wave drives micro-deformations of the diaphragm, which subsequently induces periodic variations in the optical path length of the F-P cavity, characterized by the modulation signal of the reflected probe beam intensity. After conversion by a photodetector, the second-harmonic (2f) component is extracted from this signal by a lock-in amplifier, ultimately achieving quantitative inversion of the gas concentration.

### 2.2 Working Principle of Sensor

The photoacoustic response signal S of the MFPAS to gas molecules can be expressed as:

$$S = M \cdot C_{cell} \cdot P_0 \cdot \alpha \cdot C + S_b \tag{1}$$

where M represents the acoustic sensitivity of the diaphragm, $C_{\text{cell}}$ denotes the cavity constant characterizing the generation and localized enhancement process of the PA wave, $P_0$ is the pump laser power, $\alpha$ is the absorption coefficient of the target gas, $C$ is the gas concentration, and $S_b$ represents the system background noise.

The $Si_3N_4$ diaphragm possesses a high residual tensile stress $\sigma$ after LPCVD growth. Under this condition, the mechanical behavior of the diaphragm is dominated by the tensile stress rather than the bending stiffness, which can be equivalent to an elastic membrane model with boundary constraints. Driven by a sinusoidal acoustic pressure $p_0 e^{j\omega t}$, the transverse displacement $u(r,t)$ of the clamped-boundary membrane satisfies the following equation of motion:

$$\left( h\rho \frac{\partial^2}{\partial t^2} - h\sigma\nabla^2 \right) u(r,t) = p_0 e^{j\omega t} \tag{2}$$

where h is the diaphragm thickness and $\rho$ is the material density. Neglecting the damping effect, the analytical approximation of the equation yields the displacement amplitude at the center of the diaphragm as:

$$u(0,t) = \frac{p_0 e^{j\omega t} a^2}{4h\sigma} \tag{3}$$

where a is the effective radius of the diaphragm. It can be seen that the acoustic sensitivity of the diaphragm proportional to $M = u(0)/p_0 \propto a^2$, implying that the sensitivity scales quadratically with the diaphragm radius, rather than following the fourth-power dependence of a rigid plate. Since the lateral dimensions of the F-P cavity are much smaller than the acoustic wavelength corresponding to the operating frequency, the cavity can be evaluated as a non-resonant chamber, and the internal acoustic pressure distribution tends to be uniform. Under this approximation, when the modulation frequency of the pump beam is $\omega$, the localized acoustic pressure amplitude inside the cavity is simplified as:

$$p(\omega) \propto \alpha P_0 \frac{(\gamma-1)L}{V_c} \cdot \frac{\tau_1}{\sqrt{1+(\omega\tau_1)^2}} \cdot \frac{\omega\tau_2}{\sqrt{1+(\omega\tau_2)^2}} \tag{4}$$

where L is the cavity length, $V_c = \pi a^2 L/4$ is the cavity volume, $\gamma$ is the specific heat ratio of the gas, and $\tau_1$ and $\tau_2$ are the thermal relaxation time constant of the gas inside the cavity and the damping time constant of the gas and heat flow, respectively. As can be inferred from this equation, the intracavity acoustic pressure $p \propto V_c^{-1} \propto a^{-2}$, whereas the acoustic sensitivity of the diaphragm $M \propto a^2$. Consequently, the total photoacoustic response constituted by the product of the two yields $S \propto M \cdot p \propto a^0$, meaning that it is theoretically independent of the cavity diameter. This size-independence demonstrates that the MFPAS can achieve a gas detection sensitivity comparable to that of larger cavity systems while maintaining a small diaphragm diameter, providing a theoretical guarantee for the miniaturization of the device.

The gas channels introduced by FIB milling allow the target gas to rapidly diffuse into the cavity, but simultaneously constitute an acoustic high-pass filter. Low-frequency static pressure fluctuations are suppressed via leakage through the micro-pores, effectively reducing the impact of ambient pressure noise on the diaphragm quadrature point and enhancing the stability of the sensor in practical operating environments. By locking the probe laser wavelength to the quadrature point of the F-P interference spectrum, highly sensitive demodulation of the localized intracavity PA acoustic pressure is realized by monitoring the variation in reflected light intensity as a function of diaphragm deformation. Finally, the 2f absorption signal of the target gas is extracted and the concentration is inverted by a lock-in amplifier.

### 2.3 The Analysis of Sensor

Figure 1(c) illustrates the optical micrograph of the MFPAS tip structure. The single-mode fiber is bonded and secured to the silicon substrate using epoxy resin. The F-P microcavity formed between the $Si_3N_4$ diaphragm and the end-face of the single-mode fiber simultaneously functions as both a miniature gas cell and an optical microphone. Figure 1(d) shows the scanning electron microscope (SEM) image of the diaphragm after FIB micro-pore milling. The photoacoustic response of the MFPAS is determined by the theoretical formula. A narrow-linewidth probe beam from a tunable laser is locked to the quadrature point (Q-point) of the F-P interference spectrum, and the acoustic sensitivity M is characterized by recording the variation in reflected light intensity induced by acoustic waves. As shown in Fig. 2(a), the fringe contrast of the fabricated MFPAS interference spectrum approaches 30 dB. Figure 2(b) presents the fast Fourier transform (FFT) results of the reflection spectrum, yielding a cavity length of ~196.8 μm, which is clearly within the theoretical optimization range.

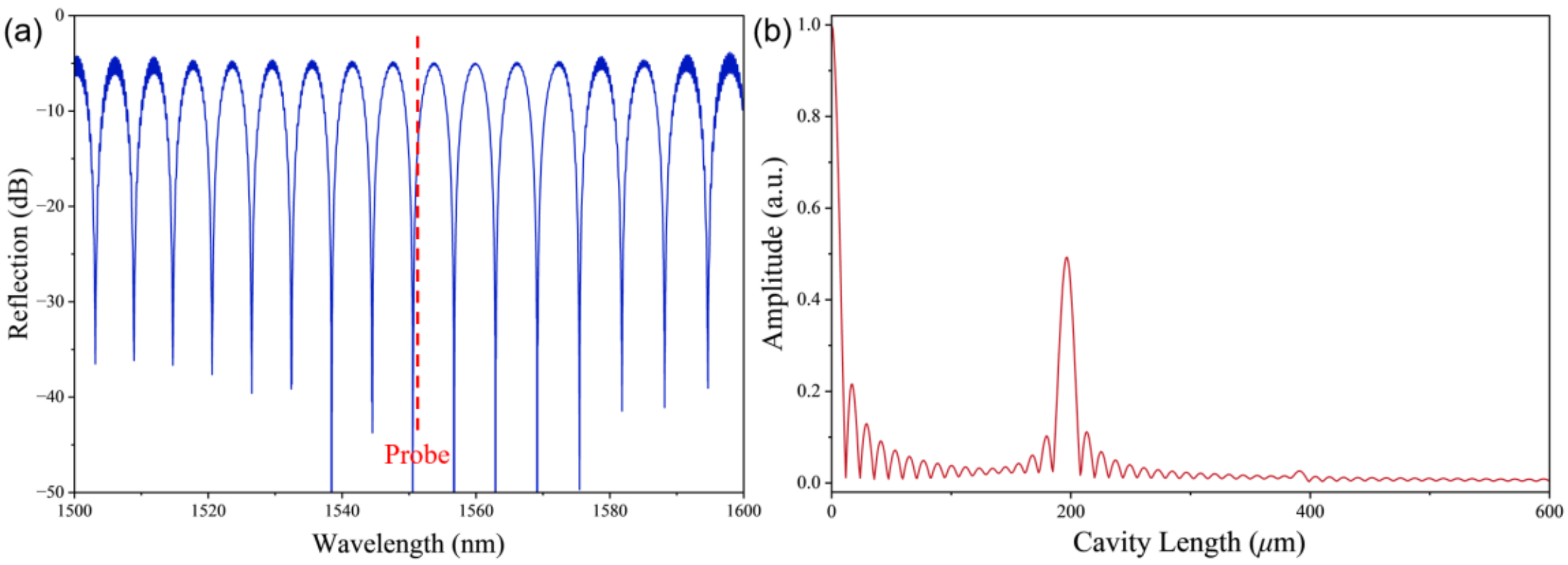

Fig. 2 (a) Interference spectrum of the MFPAS sensor. (b) FFT of the interference spectrum.

Figure 3 shows the experimental setup for MFPAS gas detection. The pump wavelength was selected as 1532.83 nm, corresponding to the P(13) absorption line of $C_2H_2$. It was generated by a distributed feedback (DFB) laser and the pump light entering the MFPAS was approximately 28 mW. After passing through a fiber isolator (ISO 1), it enters the coupler. Gas concentration measurements are conducted by combining wavelength modulation and second-harmonic (2f) detection techniques. The wavelength modulation and scanning control of the pump beam are implemented via an arbitrary waveform generator (AWG). The superimposed signal composed of the high-frequency sinusoidal modulation signal and the low-frequency sawtooth scanning signal output from the AWG is applied to the injection current driver channel of the DFB laser. Consequently, the pump wavelength is sinusoidally modulated while slowly sweeping across the P(13) absorption line. The wavelength modulation frequency is set at f = 3.25 kHz. By operating near this frequency, a high signal-to-noise ratio (SNR) for 2f detection can be achieved, and the modulation depth m is set to the optimum value of approximately 2.2 to maximize the second-harmonic (2f) signal amplitude. The probe light path is generated by a tunable semiconductor laser (TSL, wavelength around 1550 nm) and enters a 1:9 polarization-maintaining fiber coupler after passing through a fiber isolator (ISO 2). This coupler divides the probe light according to a power ratio: 90% of the power (main path) is injected via Port 1 of an optical circulator (OC) and exits from Port 2, while 10% of the power (reference path) is directly extracted to serve as the reference arm for balanced detection. After being filtered by a tunable filter (TF 1), the pump beam is combined with the main probe beam exiting Port 2 of the optical circulator via a 1×2 wavelength division multiplexing (WDM) coupler, and subsequently co-injected into the MFPAS sensing probe. The purpose of this combination scheme is to allow the pump beam to bypass the optical circulator and merge directly with the probe beam, thereby avoiding the additional insertion loss of the pump beam passing through the circulator, which effectively increases the available intracavity pump power. The reflected light from the MFPAS carries the photoacoustic modulation information. After returning through the original path of the WDM coupler, it is extracted by the optical circulator from Port 3. It then passes through a 1550 nm bandpass tunable filter (TF 2, bandwidth ±1.5 nm) to filter out the residual pump light before entering the signal arm of a balanced photodetector (BPD). The aforementioned 10% reference light simultaneously enters the reference arm of the BPD, and the two signals undergo a differential operation within the BPD. Since the reflected signal light and the reference light originate from the same TSL, both beams carry identical intensity fluctuations. After differential processing, the intensity noise is eliminated via common-mode rejection, which effectively suppresses the relative intensity noise (RIN) and low-frequency amplitude drift noise of the TSL itself, thus improving the noise-equivalent acoustic pressure detection limit of the system. The electrical signal output from the BPD is fed into a lock-in amplifier (LIA), which demodulates the 2f harmonic component using the high-frequency modulation signal from the AWG as a reference. The time constant of the LIA is set to 1 s, and the filter slope is 12 dB/Oct, corresponding to a detection bandwidth of approximately 0.1 Hz. The sensing probe is placed inside a sealed gas chamber equipped with two gas exchange ports. The $C_2H_2$ gas to be measured, with high-purity $N_2$ serving as the background gas, is continuously introduced into the gas chamber at a flow rate of 200 sccm after being mixed in a designated proportion via two mass flow controllers (MFCs) to ensure the stability and uniformity of the gas concentration.

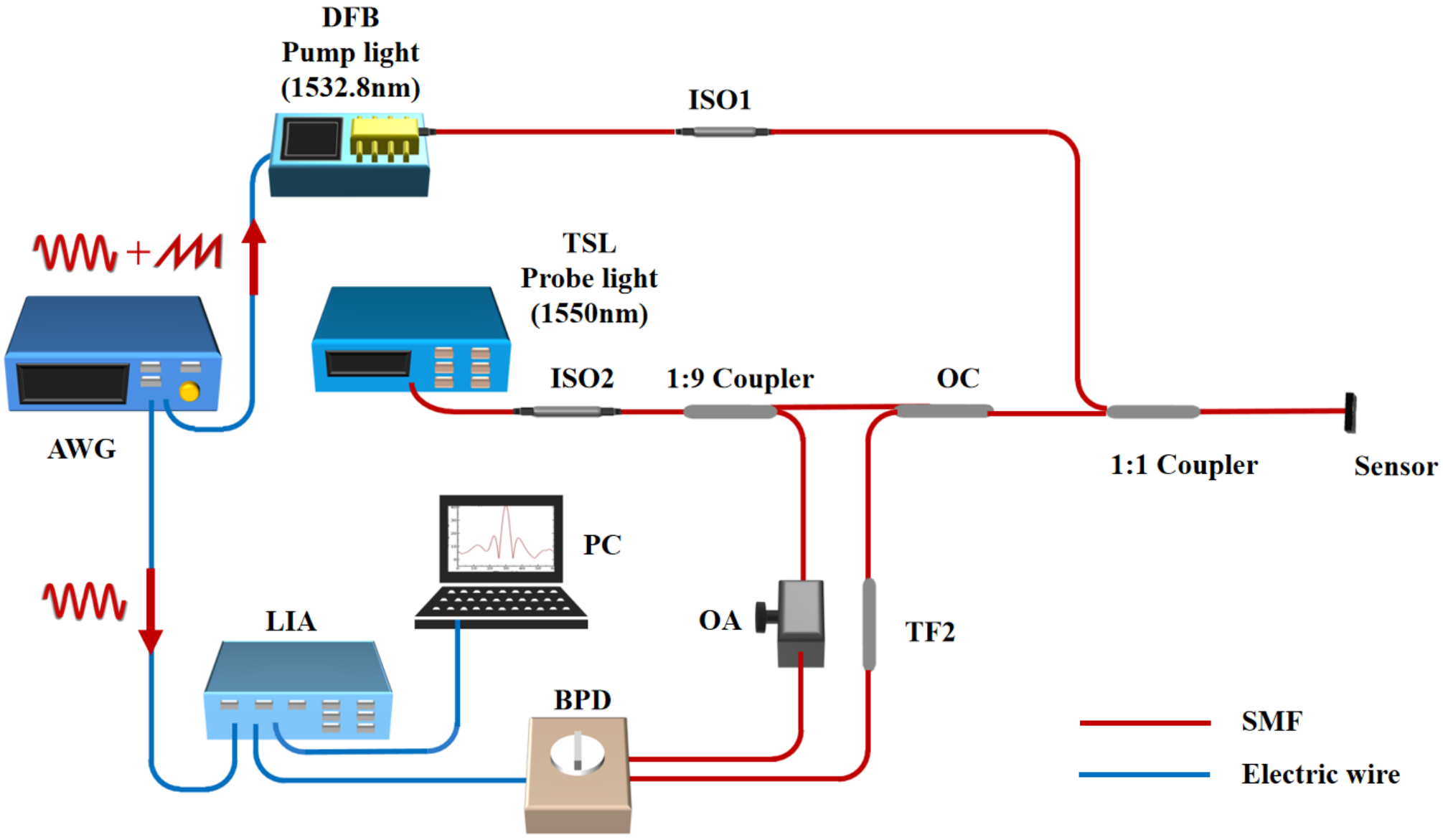

Fig. 3 The experimental setup used for testing the MFPAS gas sensor.

## 3 Results and Discussion

Under ambient conditions (room temperature and atmospheric pressure), various concentrations of $C_2H_2$ ranging widely from 1 ppm to 5000 ppm were sequentially introduced into the gas chamber for detection. The pump power entering the MFPAS was approximately 28 mW during the experiment. Figure 4(a) displays the 2f signal waveforms at different concentrations, with the inset in the upper-right corner providing an enlarged view of the low-concentration range. It is clear that even within the low-concentration regime, the signals retain distinct and well-defined photoacoustic characteristic absorption profiles. The peak-to-peak values of the 2f signals corresponding to each concentration were logarithmically fitted against the gas concentration, as shown in Fig. 4(b). The results demonstrate an excellent linear relationship between the peak-to-peak values and $C_2H_2$ concentration across the entire concentration range, with a fitting correlation coefficient of $R^2 > 0.99$. This confirms that the system satisfies the weak absorption approximation condition of the Beer–Lambert law over a broad concentration range.

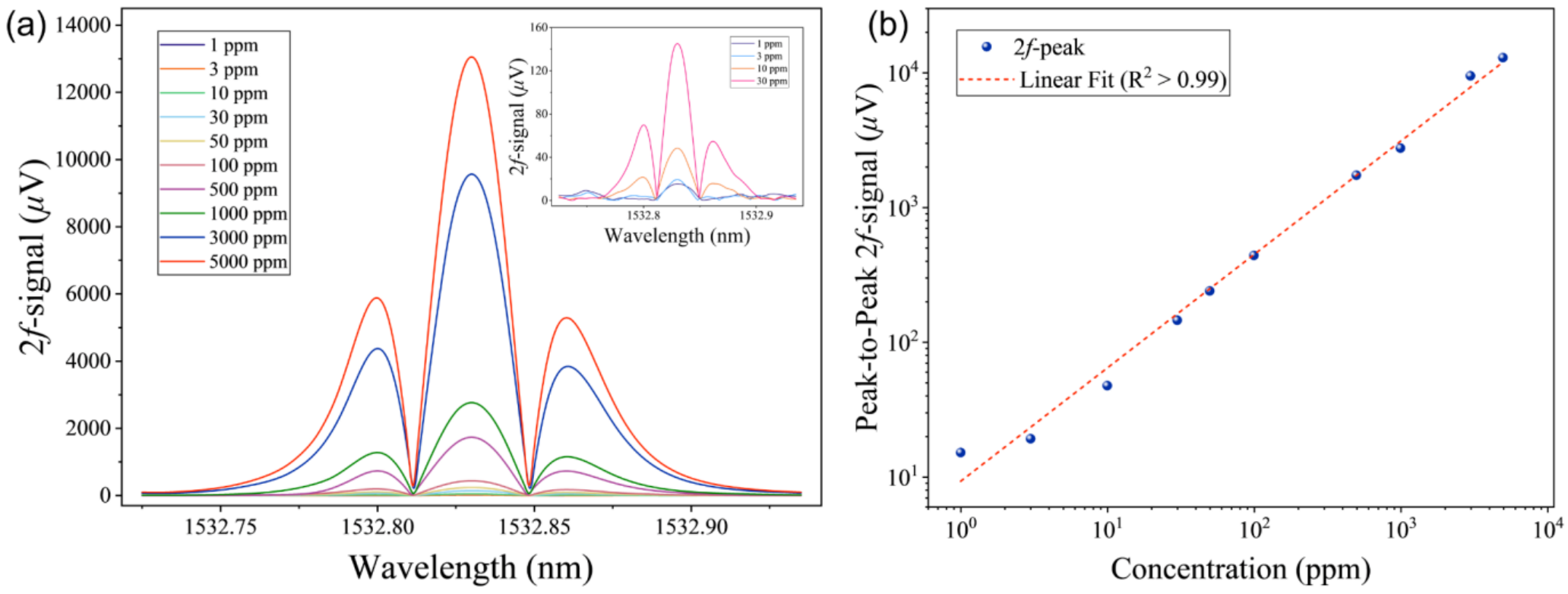

Fig. 4 (a) 2f signals of the MFPAS over a gas concentration range of 1–5000 ppm. (b) Linearity fitting results of the MFPAS.

To quantitatively evaluate the detection limit of the system, 2f signal acquisition and noise analysis were performed at a $C_2H_2$ concentration of 1 ppm, with the lock-in integration time set to 1 s. As illustrated in Fig. 5(a), the peak-to-peak value of the 2f signal at this concentration is approximately 15.4 μV, and the system 1σ noise is about 0.9 μV, yielding a signal-to-noise ratio (SNR) of ~17.14. This corresponds to an NEC of ~58.5 ppb. Considering a pump power of ~28 mW and an equivalent noise bandwidth of 0.125 Hz, the normalized noise-equivalent absorption (NNEA) coefficient is estimated to be ~$4.87 \times 10^{-9}$ $W \cdot cm^{-1} \cdot Hz^{-1/2}$. To further determine the ultimate NEC of the sensor for $C_2H_2$ detection, the system noise was continuously recorded for 12,000 s under a pure $N_2$ background atmosphere, based on which an Allan–Werle variance analysis was conducted. Figure 5(b) shows the NEC as a function of the integration time. The NEC continually decreases with increasing integration time, adhering to the $t^{-1/2}$ relationship dominated by white noise. The minimum detection limit of 25 ppb $C_2H_2$ is achieved at an integration time of ~127 s. Under the same pump power and detection bandwidth, this corresponds to an NNEA coefficient of approximately $2.08\times 10^{-9}$ $W \cdot cm^{-1} \cdot Hz^{-1/2}$. As the integration time is further extended, the NEC begins to rise, indicating that the long-term drift noise of the system gradually becomes dominant.

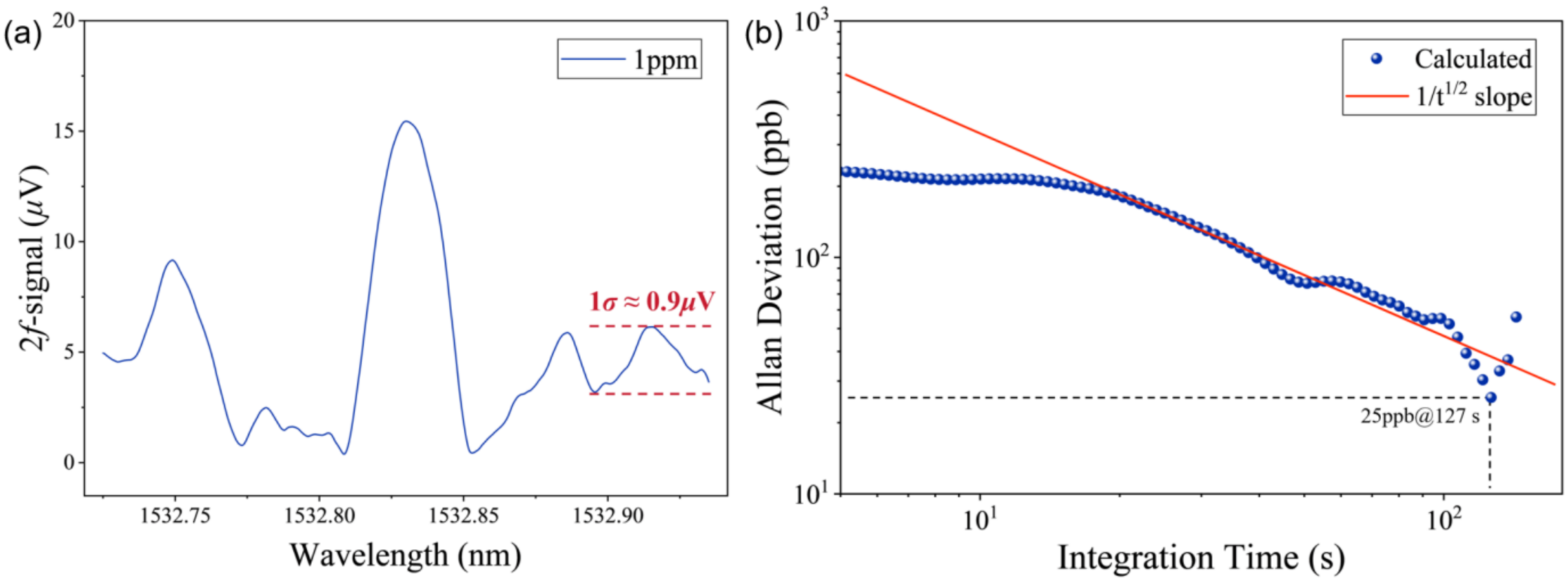

Fig. 5 (a) 2f signal analysis. (b)Allan–Werle variance analysis

The response time of the sensor is a key metric for evaluating its real-time monitoring capability. By locking the pump wavelength to the center of the $C_2H_2$ P(13) absorption line, different $C_2H_2$ concentrations ranging from 0 to 5 ppm (with pure $N_2$ as the background gas) were sequentially introduced into the gas chamber at a flow rate of 200 sccm. Concurrently, the real-time variation of the 2f signal was continuously recorded to characterize the dynamic response characteristics of the sensor, as shown in Fig. 6. As the gas concentration increases in steps of 1 ppm, the signal amplitude exhibits a distinct step-like ascent, with minimal signal fluctuations during the plateau phase at each concentration. Tests indicate that the T90 response time of the system under different concentration steps ranges between 6 and 8 seconds. This rapid response feature indicates high gas exchange efficiency inside the FPAS photoacoustic cavity. Furthermore, the system demonstrates excellent resolution for the 1-ppm concentration gradient, and no obvious baseline drift is observed during each steady-state phase, confirming that the sensor possesses outstanding short-term stability and a high SNR, making it reliable for real-time, continuous online monitoring of trace gases. It should be noted that the measured response time is primarily limited by the volume effect of the external gas chamber and the gas replacement rate. If a gas chamber with a smaller volume is adopted or the sensing probe is directly exposed to the target gas environment, the response time is expected to be further shortened to the sub-second range.

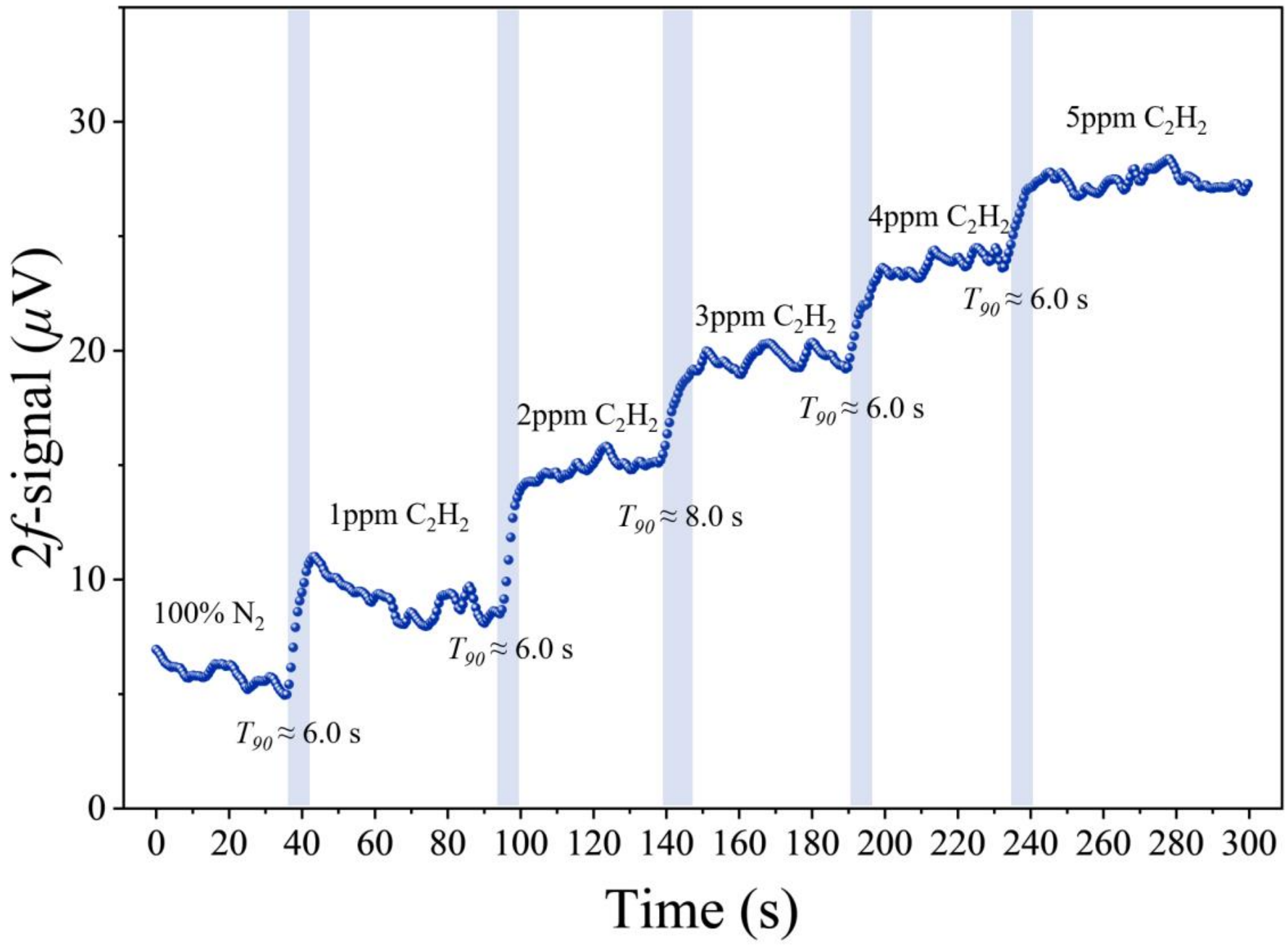

Fig. 6 Response of the MFPAS under a gas-phase acetylene concentration gradient of 0–5 ppm.

Benefiting from the ultralow chamber volume of approximately 1.5 nL, the MFPAS is capable of precise detection in extremely small gas environments. For dissolved gas detection in transformer oil, the integration of an oil–gas separation structure enables the MFPAS to perform in situ detection of dissolved gases in oil. In this work, a PF245 perfluoropolymer-based gas-permeable membrane was introduced as the oil–gas separation element. This membrane exhibits high gas permeability to typical fault-related gases such as $C_2H_2$, $CH_4$, $H_2$, and CO, while effectively preventing the penetration of liquid oil. Considering both the size of the MFPAS and the gas permeation efficiency, a 40 μm-thick PF245 membrane was selected as the encapsulation carrier to fabricate the MFPAS in-oil probe.The probe was immersed in oil samples with $C_2H_2$ concentrations of 2.78 ppm, 5.3 ppm, 6.5 ppm, 19.06 ppm, and 32.22 ppm, which were calibrated by gas chromatography. Figure 7(a) shows the 2f signal waveforms obtained at different concentrations, while Figure 7(b) presents the linear fitting relationship between the peak-to-peak value of the 2f signal and the $C_2H_2$ concentration. The results show good linearity, with a sensitivity of 1.18 μV/ppm and a correlation coefficient $R^2$ of approximately 0.99.

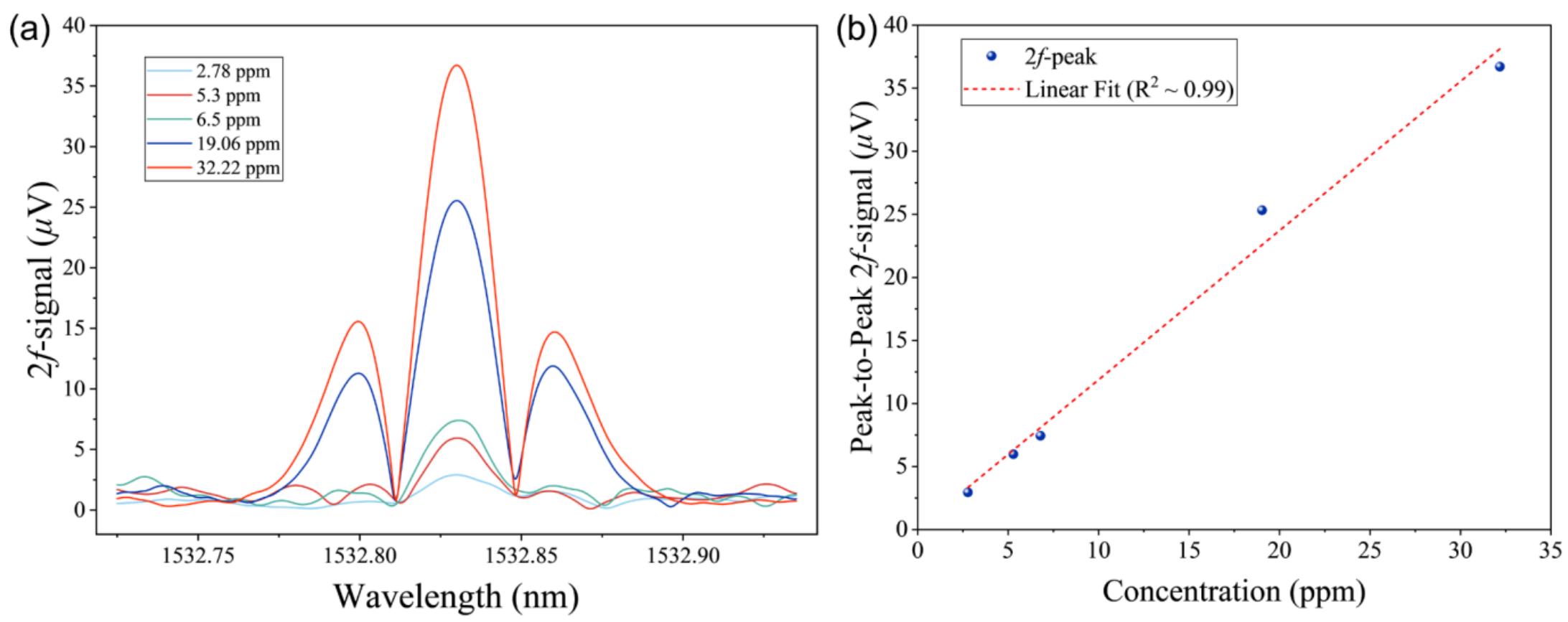

Fig. 7 (a) 2f signal at 2.4 ppm. (b) Responses under different $C_2H_2$ concentrations.

To obtain the detection limit of acetylene in oil, the response under the 2.78-ppm acetylene-in-oil sample was analyzed, as shown in Fig. 8(a). The sensor obtains a symmetrical 2f signal with a peak value of 2.9 μV at the characteristic absorption line, and the baseline 1σ noise is approximately 0.24 μV. Based on this, the NEC is calculated to be ~230 ppb, corresponding to a NNEA coefficient of ~$1.91 \times 10^{-8}$ $W \cdot cm^{-1} \cdot Hz^{-1/2}$. Additionally, by locking the pump wavelength to the center of the absorption line, oil samples with acetylene concentrations of 0.11 ppm and 1.12 ppm were respectively evaluated. As illustrated in Fig. 8(b), the MFPAS oil probe exhibits a clear response and a favorable SNR even for the extremely low-concentration oil sample containing ~1.12 ppm acetylene, with a $T_{90}$ response time of approximately 320 s. This demonstrates that the photoacoustic sensing unit not only possesses superior intrinsic detection performance in the gas phase but also maintains an outstanding trace detection capability in oil after gas-permeable, oil-blocking membrane encapsulation. Furthermore, the overall dynamic response time of the system is primarily constrained by the physical transmembrane diffusion and exchange rate of dissolved gases driven by the concentration gradient across both sides of the membrane, rather than the acoustic response limit of the photoacoustic sensing unit itself. The detection period of 320 s is significantly shorter than the hour-scale response of conventional degassing apparatuses, validating that the MFPAS oil probe combines a low detection limit with excellent timeliness for continuous online monitoring of transformer oil.

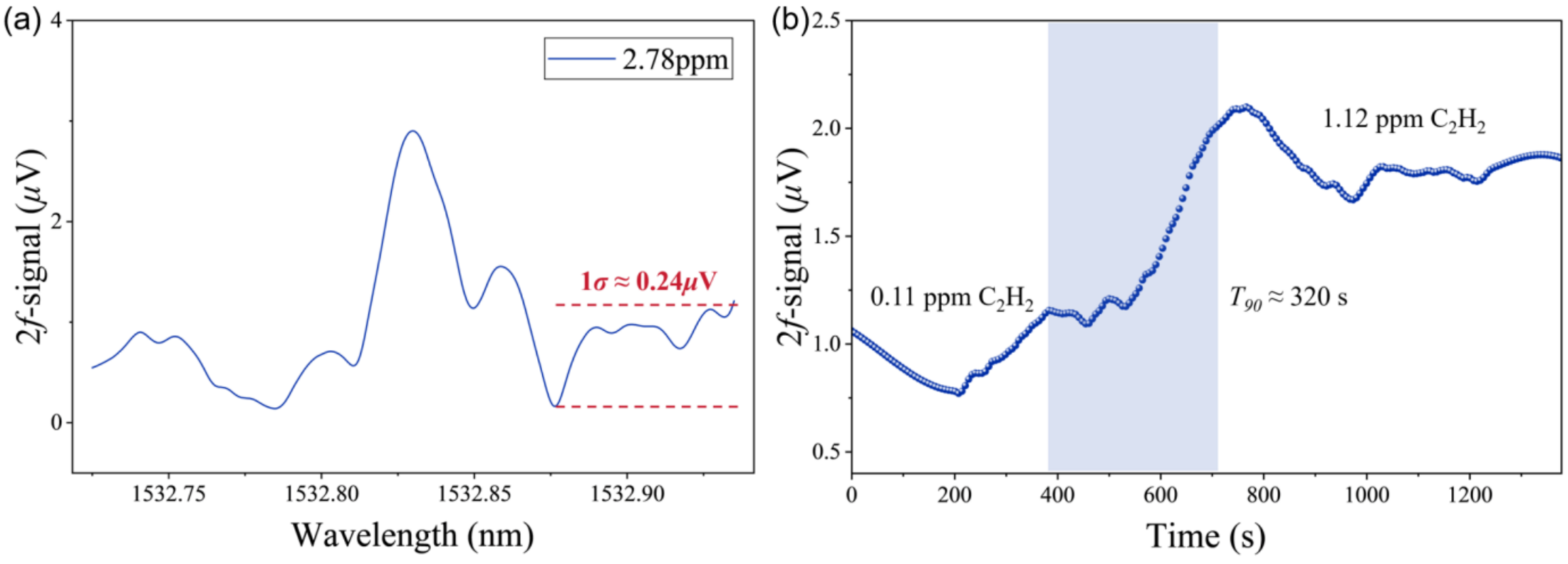

Fig. 8 (a) 2f signal analysis. (b) Response time test of the acetylene-in-oil probe in the oil phase.

Finally, the performance of this work is compared with advanced miniature fiber-optic gas sensors reported in recent years, and the results are summarized in Table 1. Considering comprehensive indicators such as sensitivity, response speed, detection volume, and fabrication consistency, the MFPAS exhibits significant overall advantages in the field of miniaturized, high-performance gas detection. Notably, the efficient and controllable introduction of gas into the microcavity remains a common challenge faced by miniaturized fiber-optic spectroscopic gas sensors. Existing schemes either rely on femtosecond laser drilling of micro-holes on the sidewalls of hollow-core fibers [11,17], utilize the natural gaps at fiber splice joints [15], leverage the inherent clearances between 3D micro-printed microstructures [13], or achieve gas exchange through the open end of a capillary. Consequently, the geometric parameters of their gas channels are difficult to regulate independently and accurately while maintaining acoustic performance. This work combines wafer-scale MEMS diaphragm micromachining with FIB milling processes for the first time, proposing a brand-new solution path. The MEMS process achieves precise and consistent fabrication of silicon nitride acoustically sensitive diaphragms at the wafer level, laying the foundation for batch manufacturing. Concurrently, by virtue of maskless direct writing and sub-micron positioning accuracy, the FIB technique realizes flexible and precise milling of micro-pore arrays on the diaphragm without introducing extra residual stress, thereby ensuring the integrity and flatness of the diaphragm. This fabrication paradigm fundamentally eliminates the inherent limitation where diaphragm-type optical microphones cannot undergo gas exchange due to their sealed structures. As a result, it endows the device with photoacoustic spectroscopic gas detection functionality while maintaining its acoustic sensitivity, opening up a completely new design paradigm for transforming existing highly sensitive diaphragm-type fiber-optic acoustic sensors into miniature photoacoustic spectrometers.

Table 1. Performance comparison with representative compact photoacoustic and photothermal gas sensors.

| **Method** | **Sensing Unit** | **NEC (ppb)** | **NNEA ($cm^{-1}$ W $Hz^{-1/2}$)** | **Response (s)** | **Dimension** |
|---|---|---|---|---|---|
| QEPAS[9] | Four-prong QTF | 390 @1 s | ~$10^{-8}$ | ~0.2 | Inner diameter 0.9 mm, length 20 mm |
| PTI[12] | 5.5 cm AR-HCF | 2.3 @670s | 2.3 × $10^{-9}$ | < 52 | ~5.5 cm（long） |
| 3D-printed PTS [13] | 0.066 mm air cavity | 160 @433 s | ~$10^{-8}$ | <0.5 | mm-scale tip |

| Method | Sensing Unit | NEC (ppb) | NNEA ($cm^{-1}$ W $Hz^{-1/2}$) | Response (s) | Dimension |
|---|---|---|---|---|---|
| MEMS FPAS (This work) | 200μm Si micro-chamber | 25 @127 s | 2.08 ×$10^{-9}$ | 6 | 3×3×0.2 $mm^3$ |

## 4 Conclusion

In response to the practical application requirements for miniaturized and highly sensitive trace gas sensing in confined spaces, a MFPAS is proposed and experimentally validated in this paper. The sensor directly butt-couples a SMF to a MEMS chip carrying a 100-nm-thick LPCVD silicon nitride diaphragm, constructing a monolithic miniature F–P photoacoustic confinement cavity with a volume of only ~1.5 nL. This study combines wafer-scale MEMS diaphragm micromachining processes with FIB technology to precisely mill micro-pores at the periphery of the suspended diaphragm, thereby overcoming the inherent limitation of conventional sealed-diaphragm optical microphones that cannot implement gas exchange. While providing highly efficient gas diffusion channels, the micro-pore array simultaneously functions as an acoustic high-pass filter. This configuration effectively suppresses low-frequency ambient pressure noise, enabling the system to stabilize its interference quadrature point without active servo control. Experimental results reveal that in gas-phase acetylene detection at 1532.83 nm, the sensor achieves a NEC of 58.5 ppb at a 1-s integration time, with a system response time of only 6 s. Allan–Werle deviation analysis further demonstrates that the NEC reaches a minimum of 25 ppb at an integration time of 127 s, corresponding to a NNEA coefficient of $2.08\times10^{-9}$ $W\cdot cm^{-1}\cdot Hz^{-1/2}$. Furthermore, after adaptive encapsulation combined with a perfluoropolymer gas-permeable membrane, the device possesses the capability for in situ detection of dissolved gases in transformer oil, yielding an NEC of 230 ppb at a 1-s integration time in the oil phase with a $T_{90}$ dynamic response time of approximately 320 s.In summary, the MFPAS presented in this study combines the advantages of a nanoliter-scale detection volume, ppb-level detection limit, low NNEA, rapid response, and wafer-scale batch fabrication consistency, exhibiting substantial engineering application potential in micro-scale detection fields such as online status monitoring of power equipment and in situ health diagnostics.